\documentclass[aps,pra,preprint,superscriptaddress,longbibliography,12pt]{revtex4-1}
\selectfont
\usepackage{setspace}
\usepackage{fancyhdr}
 \fancypagestyle{plain}{
}
\usepackage{amsmath}
\usepackage{booktabs}
\usepackage{siunitx}
\usepackage{graphicx}
\graphicspath{{FIGURE/}}
\usepackage{array}
\usepackage{verbatim}
\usepackage{graphicx}
\usepackage{dcolumn}
\usepackage{bm}
\usepackage[utf8x]{inputenc}
\usepackage{amssymb}
\usepackage{subfigure}
\usepackage{pict2e}
\usepackage[]{color}
\usepackage{wasysym}
\usepackage{mathtools}
\usepackage{hyperref}
\hypersetup{
    colorlinks = true,
    urlcolor   = blue,
    citecolor  = blue,
}
\usepackage{hyphenat}
\usepackage{titlesec}

\DeclareMathOperator{\ra}{\!\mathit{Ra}}

\newcommand*\diff{\mathop{}\!\mathrm{d}}

\usepackage{mathrsfs}

\begin{document}

\title{Experimental assessment of mixing layer scaling laws in Rayleigh-Taylor instability}

\author{Marco De Paoli}
\email[Author to whom correspondence should be addressed: ]{marco.de.paoli@tuwien.ac.at}
\affiliation{Institute of Fluid Mechanics and Heat Transfer, TU Wien, 1060 Vienna, Austria}
\affiliation{Physics of Fluids Group, University of Twente, 7500AE Enschede, The Netherlands}

\author{Diego Perissutti}
\affiliation{Polytechnic Department, University of Udine, 33100 Udine, Italy}

\author{Cristian Marchioli}
\affiliation{Polytechnic Department, University of Udine, 33100 Udine, Italy}

\author{Alfredo Soldati}
\affiliation{Institute of Fluid Mechanics and Heat Transfer, TU Wien, 1060 Vienna, Austria}
\affiliation{Polytechnic Department, University of Udine, 33100 Udine, Italy}

\date{\today}

\begin{abstract}
We assess experimentally the scaling laws that characterize the mixing region produced by the Rayleigh-Taylor instability in a confined porous medium.
In particular, we wish to assess experimentally the existence of a superlinear scaling for the growth of the mixing region, which was observed in recent two-dimensional simulations. 
To this purpose, we use a Hele\hyp{}Shaw cell. 
The flow configuration consists of a heavy fluid layer overlying a lighter fluid layer, initially separated by a horizontal, flat interface.
When small perturbations of concentration and velocity fields occur at the interface, convective mixing is eventually produced: Perturbations grow and evolve into large finger-like convective structures that control the transition from the initial diffusion-dominated phase of the flow to the subsequent
convection-dominated phase. 
As the flow evolves, diffusion acts to reduce local concentration gradients across the interface of the fingers. 
When the gradients become sufficiently small, the system attains a stably-stratified state and diffusion is again the dominant mixing mechanisms. 
We employ an optical method to obtain high-resolution measurements of the density fields and we perform experiments for values of the Rayleigh-Darcy number (i.e., the ratio between convection and diffusion) sufficiently large to exhibit all the flow phases
just described, which we characterize via the mixing length, a measure of the extension of the mixing region.
We are able to confirm that the growth of the mixing length during the convection-dominated phase follows the superlinear scaling predicted by previous simulations.
\end{abstract}

\maketitle

\section{Introduction}\label{sec:rt_intro}
When two miscible fluids with different densities move under the action of gravity the relative acceleration between the fluids generates an instability at the separation interface.
This instability, called Rayleigh-Taylor instability \citep{rayleigh1883investigation,taylor1950instability}, grows in time and leads to the formation of convective flow structures of heavy fluid moving downward and of lighter fluid penetrating upward. 
The dynamics of the flow is first controlled by diffusion, which is responsible for the thickening of the interface, initially flat and horizontal. 
Diffusion favours mixing by increasing the length of the separation interface and gives rise to a buoyancy-driven flow by inducing local gradients of concentration.
Afterwards, when the flow evolution is controlled by convection, instabilities grow and merge into large and more stable structures.
The flow persists until a stable configuration is achieved in which the density distribution is ultimately uniform over the entire domain. 
Because of its relevance in applications characterized by buoyancy-driven flows, the Rayleigh-Taylor mixing process has been widely investigated when the instability leads to a turbulent flow dominated by convection \citep{boffetta2017incompressible}.
In contrast, the evolution of the process within porous domains is far less explored \citep[see][ and references therein]{depaoli2019prf,Boffetta2020,boffetta2022dimensional} in spite of its paramount importance in many geophysical and industrial situations, including water contamination \citep{leblanc1984sewage,molen1988}, ice formation \citep{feltham2006,wettlaufer_worster_huppert_1997}, salinity inversion \citep{land1991,morton1987regional}, petroleum migration \citep{simmons2001variable} and carbon sequestration \citep{huppert2014fluid,Emami-Meybodi2015,depaoli2021influence,brouzet2022co,croccolo2022}.

In porous media flows, viscosity dominates over inertia at the pore scale and the momentum transport can be described by the Darcy law \citep{hewitt2020vigorous}. 
In this limit, a wavy and evolving interface leads to the formation of characteristic vertical, elongated structures that are usually called {\it fingers}.
The lateral spreading (hence, the width or wavelength) and the vertical growth (hence, the amplitude) of these fingers is controlled by the interplay between diffusion and convection. 
The resulting time evolution of the fingers can be characterized by the {\it mixing length}, which is defined as the tip-to-rear finger length, and provides a macroscopic measure of the vertical extension of the mixing region.
The evolution of porous Rayleigh-Taylor systems has been investigated with solutes (or chemicals) having different properties.
With respect to the properties of the species involved, possible flow configurations can be grouped in three main categories: presence of one chemical, presence of two chemicals with different diffusion coefficients, and presence two species that can chemically react \citep{de2020chemo}.
With the aid of numerical simulations and experiments, \citet{lemaigre2013asymmetric} observed that in absence of chemical reactions the mixing region grows symmetrically with respect to the initial position of the interface: The growth occurs at a velocity that depends on the nature of the fluids involved, i.e. it is controlled by the diffusivity of the species and by the initial density difference \citep{gopalakrishnan2021scalings}. 
The evolution of the system is different in presence of chemical reactions and moderate density contrast. The flow pattern develops in an asymmetric fashion \citep{lemaigre2013asymmetric}, and chemical reactions act to stabilize the flow when the product is lighter than the reactants \citep{budroni2014chemical}: A local decrease of density induced by the fact that the product is less dense creates a non-monotonic density profile \citep{bratsun2015concentration} with a minimum that hinders the penetration of the fingers into the bulk of the unmixed region. 
In contrast, a destabilizing effect can be obtained when the product provides a sufficiently larger density increase with respect to both reactants \citep{de2016chemo}.
Chemical reactions can also induce secondary instabilities in time \citep{de2020chemo}, for instance, when a sufficient amount of chemical reaction product triggers the fingered sinking of denser in the less dense reactant \citep{almarcha2011convective}.
In summary, reactions can stabilize or destabilize convection, but in all cases, they increase the mixing rate within the host phase.

In this work, we will refer to the simplest cases, consisting of one chemical specie, i.e. absence of double diffusion and chemical reactions, which produces a symmetric growth of the mixing region.
According to theoretical scaling arguments, the mixing length (that quantifies the extension of the mixing region) is expected to evolve linearly in time, i.e. with scaling exponent equal to 1, as a result of the balance between buoyancy-induced convection and dissipation due to diffusion and viscosity.
However, recent two-dimensional simulations of Darcy flows in porous media \citep{depaoli2019prf,Boffetta2020} have shown that the growth of the mixing length is superlinear, with scaling exponent equal to 1.2.

While a simplified phenomenological model to estimate the amount of mixing induced by a Rayleigh-Taylor instability has been proposed recently \citep{depaoli2019prf}, the reasons for the superlinear mixing length growth, not observed in the three-dimensional case \citep{Boffetta2020}, are yet unclear. 
In an effort to shed light on these reasons, in this work we provide a first experimental assessment of the superlinear growth, based on accurate measurements of the scaling exponent.
We use a Hele\hyp{}Shaw apparatus, which consists of two parallel transparent plates separated by a narrow gap. 
The density difference between the fluid layers is induced by the presence of a solute.
For sufficiently-small fluid velocity within the gap, a quasi-two-dimensional Poiseuille flow establishes between the plates.
This flow is known to approximate closely the Darcy-type flow observed in ideal porous media when the velocity at which the solute diffuses in the direction normal to the plates is much higher than the vertical advective velocity \citep{letelier2019perturbative,depaoli2020jfm}.

Since the pioneering work of \citet{saffman1958penetration}, the Hele\hyp{}Shaw apparatus has been widely used to mimic the behaviour of two-dimensional Darcy flows. 
However, time-dependent and highly-resolved measurements are still challenging in the Rayleigh-Taylor configuration.
This is mainly due to the experimental limitations associated with the initialization of the system, which requires to start the flow from a flat horizontal interface between the fluids.
The experimental measurements presented in this work have been performed taking care to achieve proper initial condition and minimizing the local perturbations of concentration and velocity fields, which eventually lead to mixing. 
At the same time, measurements have been performed with a high resolution in space and time, to ensure an accurate reconstruction of the concentration field, and in domains that are relatively large compared to previous literature results and extend the measurements range.
The fluids and the cell have been designed to allow a direct comparison with previous numerical simulations \citep{depaoli2019prf,depaoli2019universal}.
We analyzed the evolution of the mixing length and the wavenumber power spectra (i.e., the finger size).
Concerning the mixing length, we observe the same superlinear behaviour predicted by the simulations \citep{depaoli2019prf,Boffetta2020}, and we provide an explanation for this finding in terms of finite size effects of the flow domain.
With respect to the fingers evolution, we find again that the measured number of fingers is in very good agreement with that predicted numerically \citep{depaoli2019universal}.
Overall, we are able to assess successfully the scaling relations provided by numerical simulations. We believe that this result can open new perspectives for future modelling and parameterization of convective flow in confined porous media. 

This paper is organised as follows.
In Sec.~\ref{sec:rt_method}, we formulate the problem, describe the experimental setup and recall the assumptions underlying the referenced numerical simulations.
The results are presented in Sec.~\ref{sec:rt_resul} both in qualitative (flow phenomenology) and quantitative (mixing length and wavenumbers) terms.
Finally, conclusions are discussed in Sec.~\ref{sec:rt_concl}.

\section{Methodology}\label{sec:rt_method}
Our experiments were specifically designed to mimic the mixing process
produced by the Rayleigh-Taylor instability in a two-dimensional, saturated and confined porous medium. In addition, they were carried out imposing a linear dependency of density and solute concentration, and ensuring the occurrence of a Darcy-type flow to allow direct comparison with previous numerical simulations \citep{depaoli2019prf,depaoli2019universal}.

\subsection{Problem formulation }\label{sec:rt_num}
\begin{figure}
\centering
\includegraphics[width=0.95\columnwidth]{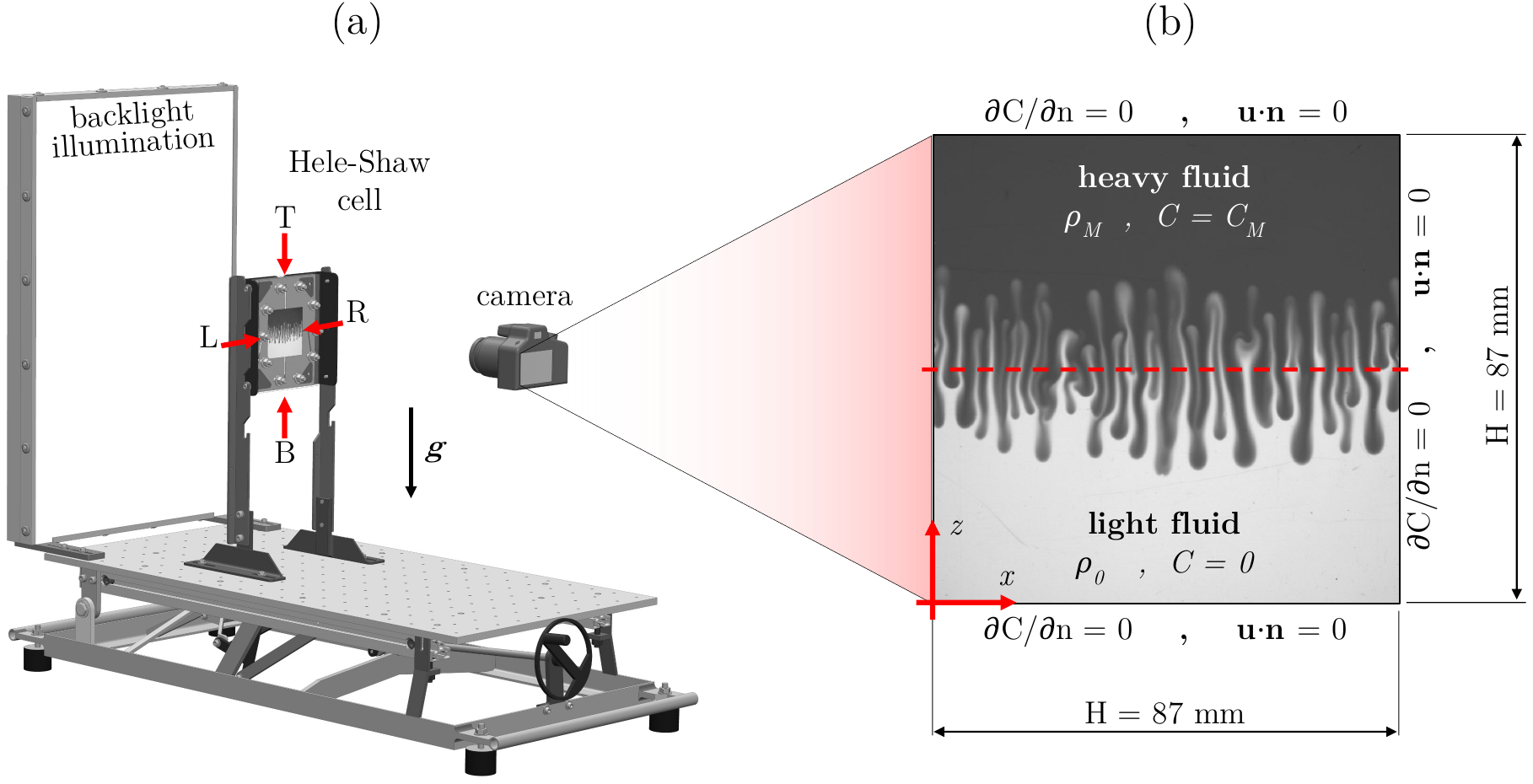}
\caption{\label{fig:sk} 
(a) Experimental setup. 
The Hele\hyp{}Shaw cell is shown, with explicit indication of the position of the connection gates (top, bottom, left and right, indicated as T, B, L, and R respectively), sCMOS camera and backlight illumination.
Fluids are supplied to the cell through a system (sketched in Fig.~\ref{fig:circ}) consisting of a pump, pipes and valves.
The cell is aligned in the vertical direction, namely parallel to the acceleration due to gravity, $\mathbf{g}$.
(b) Sketch of the experimental domain with explicit indication of the boundary conditions (no flux of mass or solute through the walls).
The reference frame ($x,z$) as well as the initial position of the interface (red dashed line) are indicated, with the heavy fluid (density $\rho_M$, concentration $C=C_M$) initially lying on top of the lighter fluid ($\rho_0$, $C=0$).
The background field consists of one of the images collected, and provides a qualitative picture of the flow for the present physical configuration.
} 
\end{figure}

The process of convective dissolution is studied via the Rayleigh-Taylor instability, corresponding to two layers of fluid of different density, initially in an unstable configuration and subject to relative acceleration under the action of gravity \citep{boffetta2017incompressible}. 
The process is simulated in the frame of porous media flows, mimicked with the aid of a Hele\hyp{}Shaw cell, i.e. two parallel and transparent plates of height $H$ separated by a narrow gap $b$, as shown in Fig.~\ref{fig:sk}(a). 
The cell is initially saturated with two miscible fluids having same viscosity ($\mu$), arranged such that the heavy fluid (density $\rho_M$) lies on top of the lighter one (density $\rho_0$), as sketched in Fig.~\ref{fig:sk}(b).
The maximum density difference within the system is $\Delta\rho =\rho_M-\rho_0$ and is induced by the presence of a solute, namely potassium permanganate (KMnO$_4$).
The amount of solute is quantified by its concentration $C$, which is maximum at the upper layer ($C=C_M$) and minimum at the lower layer ($C=0$).

The evolution of the system is controlled by the contributions of buoyancy, which tends to bring the fluids in a stable configuration and it is associated with the buoyancy velocity $U$, and diffusion, which acts to reduce local concentration gradients and to increase the mixing in the domain, and it is quantified by the diffusion coefficient $D$.
The relative importance of these two contributions can be estimated by the Rayleigh-Darcy number, which represents the governing dimensionless parameter of the system, and is defined as:
\begin{equation}
\ra=\frac{UH}{\phi D}=\frac{H}{\ell}\text{  ,}
\label{eq:rt_ra0exp} 
\end{equation} 
with:
\begin{equation}
U=\frac{g\Delta\rho k}{\mu}\quad,\quad\ell=\frac{\phi D}{U}\quad,
\label{eq:rt_ra0exp2} 
\end{equation} 
the buoyancy velocity and the length scale over which advection and diffusion balance,
respectively \citep{slim2014solutal}.
The buoyancy velocity, is defined here as the combination of buoyancy (controlled by the acceleration due to gravity $g$, density contrast $\Delta\rho$, and medium permeability $k$) and dissipation (controlled by fluid viscosity, $\mu$).
The length scale $\ell$ is particularly important when a comparison among systems having different Rayleigh numbers is required.
Permeability $k$ and porosity $\phi$ are the characteristic properties of the porous medium mimicked by the Hele\hyp{}Shaw cell. In our experiments, $k=b^2/12$ and $\phi=1$. 
In order to prevent potential three-dimensional effects we followed the indications emerged in the theoretical paper by \citet{letelier2019perturbative} and later confirmed experimentally by \citet{depaoli2020jfm} and \citet{alipour2020concentration}. 
We introduce the anisotropy ratio, which is an additional parameter describing the geometry of the cell, defined as: 
\begin{equation}
\epsilon=\frac{b}{\sqrt{12}H}\text{  .}
\label{eq:rt_ra0exp3} 
\end{equation} 
and we can use this parameter in combination with the Rayleigh
number to ensure that the flow will be of the Darcy-type if the condition $\epsilon^2\ra\to0$ is met. 
In our study, $\epsilon\approx10^{-3}$ and $4.4 \times 10^{-3} < \epsilon^2\ra < 5.4 \times 10^{-2}$, indicating that we are very close to the theoretical limit corresponding to the Darcy-type regime.

Finally, we are interested in examining a system with boundaries that are impermeable to both fluid and solute. Hence, the following conditions are imposed along the cell walls:
\begin{equation}
\mathbf{u}\cdot\mathbf{n}=0\quad,\quad \frac{\partial C}{\partial n}=0 \text{  ,}
\label{eq:rt_ra0exp4}
\end{equation}
with $\mathbf{n}$ the unit vector perpendicular to the boundary.

\subsection{Experimental setup}\label{sec:expset}
We used a transparent Hele\hyp{}Shaw cell (Polymethyl Methacrylate - PMMA - thickness~8~mm).
The wall-normal depth of the fluid layer is constant for all the experiments performed and corresponds to the value $b=300~\mu$m. 
An impermeable rubber (Klinger-sil C-4400, thickness $300$~$\mu$m), is placed between the two acrylic plates, and is used to seal the region containing the fluid but also as a spacer between the cells surfaces.
The sealing is laser-cut so that the domain investigated is a square of side $H=87$~mm. 
On each side, the midpoints are connected to pipes that allow the cell filling/emptying process. 
The transparent sheets and the gasket are held in place by an outer metal frame and a set of 10 bolts, which are tightened with the same torque, to prevent the formation of a non-uniform film thickness. 
The experimental apparatus is represented in Fig.~\ref{fig:sk}(a). 
Backlighting is provided to the cell by a tunable LED system (150 lamps covered by a diffusing glass). 
Finally, the evolution of the flow is recorded by a digital sCMOS camera (FlowSense USB 2M-165, Dantec Dynamics). 
The imaging system and settings will be described in detail in Sec.~\ref{sec:calib}.
Next, we describe the fluids adopted and the processes of concentration reconstruction.

\begin{figure}[t!]
\centering
\includegraphics[width=0.95\columnwidth]{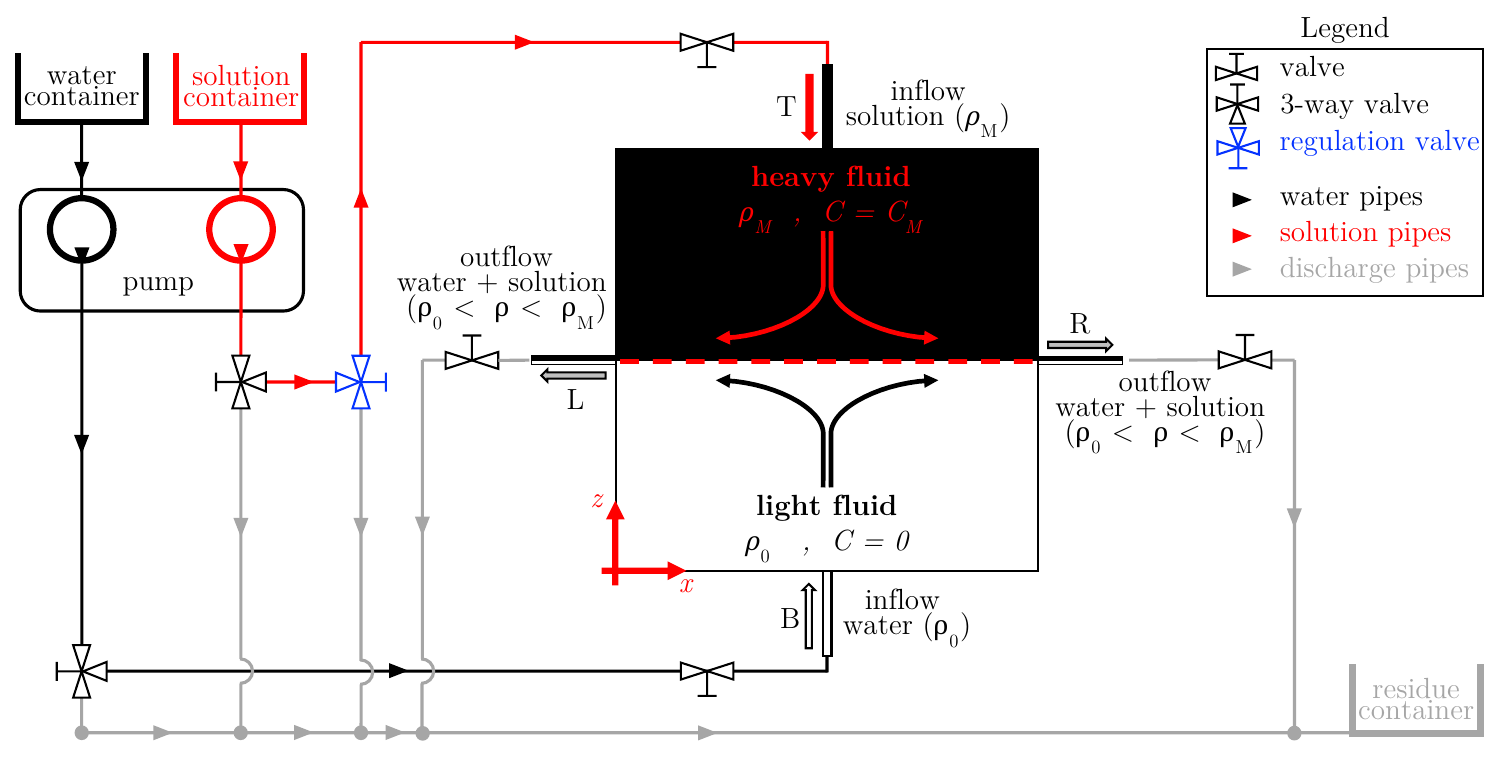}
\caption{\label{fig:circ}
Circuital scheme of the experimental setup employed to obtain the desired initial conditions \citep[][]{shi2006acceleration,shi2008novel,gopalakrishnan2017relative}.
A pumping system (peristaltic pump, Watson-Marlow 502 S) is used to inject water (density $\rho_0$) and solution ($\rho_M$) from two separate containers, through the gates B and T, respectively. 
To set up the initial condition, the valves are placed in ``loading mode'' and the exceeding solution is collected from gates L and R into a residue container.
The flow through the gates, which is used to modulate the shape of the interface between the two fluid layers, is controlled by a regulation valve.
Finally, when the initial condition is achieved, the system is set in ``experiment mode'' and the cell is by-passed by the fluids, which are channeled to the residue container. 
} 
\end{figure}

The initial condition of the experiments is an unstable configuration in which a layer of heavy fluid sits on top of a layer of lighter fluid, these layers being initially unmixed.  
To realize this challenging configuration, we employ a circuit made of a pump, valves and pipes, represented in Fig.~\ref{fig:circ}. 
This approach has already been used in previous studies \citep[see ][]{shi2006acceleration,shi2008novel,gopalakrishnan2017relative}.
The entire system is initially filled with degassed water, to minimize the formation of bubbles.
Then the cell is filled with the working fluids, which are driven by a peristaltic pump (Watson-Marlow 502 S) with the valves in ``loading mode''.
In this configuration, the pump is able to inject degassed water and a homogeneous solution of KMnO$_4$ (concentration $C_M$) from two separated containers into the cell. 
Specifically, the solution and the water are injected from the top and bottom gates of the cell, indicated with (T) and (B) in Fig.~\ref{fig:circ}. 
In this configuration, the valves connected to the left and right gates of the cell (L and R, in Fig.~\ref{fig:circ}) are both open and allow the mixture of water and solution to leave the cell and be disposed in a residue container.
Therefore, a flow from the top/bottom walls towards the left/right wall takes place, as indicated by the red and black arrows in Fig.~\ref{fig:circ}.
As a result, a sharp interface forms between the two liquid layers, while the vertical  concentration gradient makes the fluid-fluid diffusive interface smoother.
The competition between horizontal advection and vertical diffusion determines the interface thickness, which can be controlled by the flow rate (see Appendix~\ref{sec:appincond} for further details on the interface thickness). 
A regulation valve is employed to balance the flow rate between the top (T) and bottom (B) gates to achieve a straight interface.
When the initial condition is set, the valves are placed in ``experiment mode'', and the cell is by-passed by the inflow.
The buoyancy-driven flow takes place and image acquisition starts.

The two fluids are fully miscible but have different density. Specifically, we
used water for the lower (i.e. lighter) fluid layer and an aqueous solution of potassium permanganate (KMnO$_4$) for the upper (i.e. denser) fluid layer.
While water density is nearly constant among the experiments (it is only dependent on temperature), the density of the KMnO$_4$ solution can be varied by changing the solute concentration. 
We consider the coordinate system sketched in Fig.~\ref{fig:sk}(b), where $x$ and $z$ are the spatial coordinates in horizontal and vertical direction, respectively.
In this reference frame, the initial condition can be expressed in terms of the solute concentration as follows: $C(x,z\le H/2,t=0)=0$ (lower half of the cell) and $C(x,z\le H/2,t=0)=C_{M}$ (upper half of the cell), with $C_M$ the initial concentration of the KMnO$_4$ solution (upper fluid layer).
We consider that the dynamic viscosity, $\mu=9.2\times10^{-4}$~Pa$\cdot$s, is constant and independent of the solute concentration \citep{slim2013dissolution}.
Similarly, we assume that the diffusion coefficient is not sensibly affected by either
the solute concentration or the local values of velocity $D=1.65\times10^{-9}$ m$^{2}$/s.
This value has been measured and reported in literature, and it is in excellent agreement with theoretical predictions based on electrical neutrality of a simple salt at infinite dilution \citep[see ][ and references therein]{slim2013dissolution}.
The working fluids have been chosen because of the linear dependency of the density that characterizes the resulting solution with respect to the solute concentration.
This feature is essential to make reliable comparisons between experiments and simulations, which is one of the main objectives of the present work.

The density of an aqueous KMnO$_{4}$ solution can be written as a function of the fluid temperature, $\vartheta$, the water density at temperature $\vartheta$, $\rho_0$, and the KMnO$_{4}$ concentration, $C$ \citep{Novotny1988}.
Please note that also the water density can be computed from the temperature of the fluid, and therefore the parameters that need to be measured to determine the solution density are $\vartheta$ (which we assume to be uniform over the cell and constant during the experiment) and the local value of $C$ (inferred from optical measurements).
The density difference between the two fluid layers is $\Delta\rho =\rho_{M}-\rho_0$, where $\rho_{M}=\rho(C=C_{M})$ and $C_{M}$ are determined prior to each experiment.
At a given temperature, the density of the mixture obtained from empirical correlations \citep{Novotny1988} is well described by a function of $C$ and $C^{3/2}$.
However, within the range of mass fraction considered here, this correlation is well approximated by a linear function of the solute concentration (see Appendix~\ref{sec:appfluids}) as:
\begin{equation}
\rho=\rho_M\biggl[1+\frac{\Delta\rho }{\rho_MC_M}\bigl(C-C_M\bigr)\biggr]\textit{  .}
\label{eq:rt_eq4a}
\end{equation}
This equation complies with the assumption of linear density-concentration dependency that is typically made in simulations, thus ensuring a reliable comparison.
Measurements of the concentration field are inferred from the transmitted light intensities, this process being described in detail in the next section.

\subsection{Experimental procedure  and quantification of uncertainties}\label{sec:calib}

\begin{figure}
  \centering
\includegraphics[width=0.9\columnwidth]{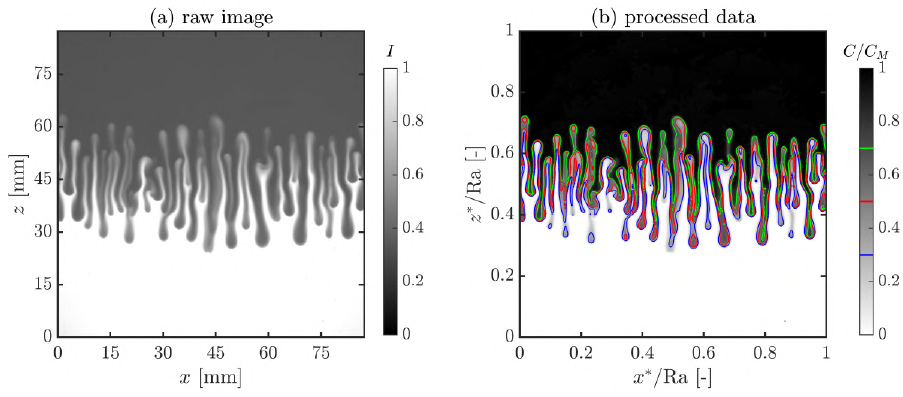}
\caption{\label{fig:rec} 
Reconstruction of the concentration field. 
(a) Raw image consists of a light intensity distribution, $I$.
(b) Upon processing to remove light intensity noises, scratches and non-uniformities of backlight illumination (see \citep{supp_info} for further details), the dimensionless concentration field is obtained. 
To visualise the interface between the two fluids, corresponding to the fingers, three iso-contours of normalised concentration are shown, corresponding to $C^*=C/C_M=0.3$ (blue), 0.5 (red) and 0.7 (green).}
\end{figure}

The results discussed in this paper are based on measurements of the time-dependent concentration field of the solute, $C(x,z,t)$, inside the Hele\hyp{}Shaw cell.
To perform these measurements, we follow the method introduced by \citet{slim2013dissolution} and later used in a number of studies \citep[][]{ching2017convective,alipour2019PAMM,alipour2020concentration,depaoli2020jfm}.
Optical measurements of light intensity distribution, $I(x,z)$, are obtained with the aid of a sCMOS camera (FlowSense USB 2M-165, Dantec Dynamics, exposure time used $1$~ms, resolution 2Mpx, frame rate 0.35~Hz to 5~Hz, depending on $\ra$).
The measurements require a calibration process, briefly described in Appendix~\ref{sec:appmeas}, in which the value of light intensity is
associated with the value of solute concentration.
The greyscale images so obtained, an example of which is reported in Fig.~\ref{fig:rec}(a), are preprocessed to reduce the effect of non-uniformities in the distribution of the light intensity over the cell, \citep{supp_info}).
Finally, since the relationship between mass fraction and solute concentration is known (see Fig.~\ref{fig:fl}(b) of Appendix~\ref{sec:appmeas}), the reconstruction of the concentration and density fields is performed. An example of this reconstruction is provided in Fig.~\ref{fig:rec}(b).

An important issue that must be assessed is the uncertainty on the measured value of the Rayleigh number, $\ra$, defined as in Eq.~\eqref{eq:rt_ra0exp}.
The uncertainty value, $\delta\ra$, is obtained from propagation-of-error analysis \citep{taylor1997introduction} and reads as:
\begin{equation}	
\frac{\delta \ra}{\ra}=
\biggl[
\left(\frac{\delta g}{g}\right)^{2}+
\left(\frac{\delta \Delta\rho }{\Delta\rho } \right)^{2}+
\left(2\frac{\delta b}{b}\right)^{2}+
\left(\frac{\delta H}{H}\right)^{2}+
\left(\frac{\delta \mu}{\mu} \right)^{2}+
\left(\frac{\delta D}{D} \right)^{2}
\biggr]^{1/2}
\text{  ,}	
\label{eq:uncer}	
\end{equation}
where $\delta \bullet$ indicates the uncertainty associated to the quantity $\bullet$ that represents each of the variables appearing in Eq.~\eqref{eq:rt_ra0exp}.
We assume that the properties $\mu$, $D$ and $g$ are estimated with constant relative uncertainty equal to $1\%$.
The uncertainty on the domain height, $\delta H$, is assumed to be 0.5~mm, primarily
due to a possible misplacement of the sealing on the cell rather than to the cutting precision.
The accuracy on the gap thickness is assumed to be $\delta b=10\mu$m. 
The uncertainties just introduced are common to all the experiments. 
In contrast, the uncertainty of the density difference $\delta\Delta\rho $
is a function of the Rayleigh number, being derived from Eq.~\eqref{eq:novot} as:
\begin{equation}
\label{formula}
    \delta\Delta\rho  = \sqrt{\frac{\left(\frac{\partial f}{\partial \vartheta }\delta \vartheta\right)^2+
    \left(\frac{\partial f}{\partial C}\rho\delta \omega\right)^2}{1-\left(\omega\frac{\partial f}{\partial C}\right)^2}},
\end{equation}
where $\delta \vartheta$ is given by the instrument (0.5$^\circ$C).
To use Eq. (\ref{formula}), the uncertainty on the concentration must be known. This is computed from Eq.~\eqref{eq:defmf} as:
\begin{equation}
    \delta C = \sqrt{\omega^2(\Delta\rho )^2+\rho^2(\delta\omega)^2}.
\end{equation}
The mass fraction, $\omega$, defined as the ratio between the mass of solute and the mass of solution, is determined measuring first the mass of water ($m_\text{w}$) and then the overall mass of solution ($m_\text{sol}$), i.e. $\omega=(m_\text{sol}-m_\text{w})/m_\text{sol}$.
As a result, the uncertainty $\delta\omega$ reads as:
\begin{equation}	
\frac{\delta \omega}{\omega}=\frac{1- \omega}{\omega}\sqrt{
\left(\frac{\delta m_\text{sol}}{m_\text{sol}} \right)^{2}+
\left(\frac{\delta m_\text{w}}{m_\text{w}} \right)^{2}
}\text{  ,}
\label{eq:uncer2}	
\end{equation}
with $\delta m_\text{w}=\delta m_\text{sol}=0.001$~g (high-precision scale, Sartorius Acculab Atilon model ATL-423-I, $\pm0.001$~g).
The list of the solution properties, flow parameters and corresponding uncertainties is reported in Tab.~\ref{tab:listex}.

\begin{table}
\setlength{\tabcolsep}{7pt}
\footnotesize
\begin{tabular}{@{} c c c c c c @{}}
\toprule
\toprule
E\# &  $\vartheta$ & $\omega_{M}$ & $C_{M}$    & $\Delta\rho$ & $\ra$ \\
     &  [$^{\circ}C$] &   & [kg/m$^{3}$] & [kg/m$^{3}$]   &  \\ \midrule
 E1  &  $24.0\pm0.5$  &  $\left(1.41\pm0.006\right)\times10^{-3}$  &  $1.427\pm0.006$  &  $1.132\pm0.002$  &  $\left(4.39\pm0.08\right)\times10^{3}$ \\ 
 E2  &  $19.0\pm0.5$  &  $\left(2.30\pm0.006\right)\times10^{-3}$  &  $2.316\pm0.006$  &  $1.805\pm0.002$  &  $\left(7.00\pm0.12\right)\times10^{3}$ \\ 
 E3  &  $21.0\pm0.5$  &  $\left(3.94\pm0.006\right)\times10^{-3}$  &  $3.949\pm0.007$  &  $3.099\pm0.004$  &  $\left(1.20\pm0.02\right)\times10^{4}$ \\ 
 E4  &  $23.5\pm0.5$  &  $\left(6.42\pm0.006\right)\times10^{-3}$  &  $6.441\pm0.008$  &  $5.101\pm0.008$  &  $\left(1.98\pm0.03\right)\times10^{4}$ \\ 
 E5  &  $21.0\pm0.5$  &  $\left(1.08\pm0.001\right)\times10^{-2}$  &  $10.824\pm0.010$  &  $8.495\pm0.013$  &  $\left(3.30\pm0.06\right)\times10^{4}$ \\ 
 E6  &  $21.0\pm0.5$  &  $\left(1.76\pm0.001\right)\times10^{-2}$  &  $17.828\pm0.015$  &  $13.996\pm0.025$  &  $\left(5.43\pm0.09\right)\times10^{4}$ \\ 
 \bottomrule
\bottomrule
\end{tabular}
\caption{\label{tab:listex}
List of flow parameters, solution properties and corresponding uncertainties.
Different realizations have been performed for all the Rayleigh numbers, $\ra$. 
Fluid temperature ($\vartheta$) is reported, as well as solution mass fraction ($\omega_M$), concentration ($C_{M}$) and corresponding density difference ($\Delta\rho $).
The dimensions of the cell are kept constant for all the experiments ($H\times H$, with $H=(87\pm0.5)$~mm, thickness $b=(300\pm10)$~$\mu$m). 
Fluid properties are discussed in Sec.~\ref{sec:expset}.}
\end{table}

\subsection{Numerical simulations and dimensionless variables}

To assess the mixing layer scaling laws, we wish to make a direct comparison of the present experimental results with direct numerical simulations of convection in two-dimensional porous media \citep{depaoli2019prf,depaoli2019universal}. We perform this comparison in the range $347\le \ra \le 19953$, for which a very good overlap of the governing parameters exists between experiments and simulations.
To make the paper self-contained, we recall next the governing equations and the numerical details.

An isotropic and homogeneous vertical porous slab with permeability $k$ and porosity $\phi$ is considered. 
The porous domain is initially saturated with two miscible fluids having same viscosity ($\mu$) but different density, arranged such that the heavy fluid (density $\rho_M$) tops the lighter one (density $\rho_0$).
This unstable configuration replicates exactly the one sketched in Fig.~\ref{fig:sk}(b). 
As done in the experiments, the density difference is induced by the presence of a solute, with maximum solute concentration $C=C_M$ in the upper layer and minimum
solute concentration $C=0$ in the lower layer.
The flow is assumed to be incompressible and controlled by the Darcy's law:
\begin{equation}
\frac{\partial u}{\partial x}+\frac{\partial w}{\partial z}=0\textit{  ,}
\label{eq:rt_dim1}
\end{equation} 
\begin{equation}
\frac{\mu}{k}u=-\frac{\partial p}{\partial x}\quad,\quad
\frac{\mu}{k}w=-\frac{\partial p}{\partial z}-\rho g\textit{  ,}
\label{eq:rt_dim2}
\end{equation} 
where $u$ and $w$ are the horizontal ($x$) and vertical ($z$) velocity components, while $p$ and $\rho$ are the local pressure and density, respectively. 
We consider that the Oberbeck-Boussinesq approximation applies: This is a reasonable assumption for geophysical problems such as geological carbon dioxide sequestration \cite{landman2007heat}.
In addition, we consider that the density of the mixture is a linear function of
the solute concentration $C$ [see Eq.~\eqref{eq:rt_eq4a}], and that the following transport equation for $C$ holds:
\begin{equation}
\phi\frac{\partial C}{\partial t}+u\frac{\partial C}{\partial x}+w\frac{\partial C}{\partial z}=\phi D \left(\frac{\partial^2 C}{\partial x^{2}}+\frac{\partial ^2 C}{\partial z^{2}}\right),
\label{eq:rt_dim3}
\end{equation}
with $t$ the time and $D$ the solute diffusivity, which we assume to be constant. 

To make Eqs.~\eqref{eq:rt_dim1}-\eqref{eq:rt_dim3} dimensionless,
we use the buoyancy velocity $U$ and the length scale $\ell$, defined as in Eq.~\eqref{eq:rt_ra0exp2}.
In addition, the following dimensionless variables are introduced \citep{depaoli2016solute,cheng2012effect,slim2014solutal}:
\begin{equation}
p^*=\frac{p}{\Delta \rho_M g \ell}\quad ,\quad
C^*=\frac{C}{C_M} \quad ,\quad
t^*=\frac{U}{\phi \ell}t \quad.
\label{eq:rt_adim2}
\end{equation}
The resulting governing equations in dimensionless form read as:
\begin{equation}
\frac{\partial u^*}{\partial x^*}+\frac{\partial w^*}{\partial z^*}=0,
\label{eq:rt_eq5a}
\end{equation}
\begin{equation}
u^*=-\frac{\partial P^*}{\partial x^*}\quad,\quad w^*=-\frac{\partial P^*}{\partial z^*}-C^*,
\label{eq:rt_eq5}
\end{equation}
\begin{equation}
\frac{\partial C^*}{\partial t^*}+u^*\frac{\partial C^*}{\partial x^*}+w^*\frac{\partial C^*}{\partial z^*}=\frac{\partial^2 C^*}{\partial x^{*2}}+\frac{\partial ^2 C^*}{\partial z^{*2}},
\label{eq:rt_eq6}
\end{equation}
where the superscript $^*$ indicates dimensionless variables, $P^*=p^*+z^*(\rho_M/\Delta\rho -1)$ is the reduced pressure and $\Delta\rho $ is the density difference between the upper and the lower fluid layers.

The controlling parameter of the present system is the Rayleigh-Darcy number $\ra$, defined as in Eq.~\eqref{eq:rt_ra0exp}.
Unlike the Hele\hyp{}Shaw experiments, in which the geometry of the cell is characterised by the anisotropy ratio, two-dimensional Darcy simulations do not need additional dimensionless parameters: The Rayleigh-Darcy number describes completely the fluid properties ($\Delta\rho ,\mu,D$), the porous medium properties ($k,\phi$) and the domain size ($H$). 
Although not explicitly appearing in the governing Eqs.~\eqref{eq:rt_eq5a}-\eqref{eq:rt_eq6}, $\ra$ enters the picture as the dimensionless height of the domain.
Impermeable boundary conditions (i.e. no-flux) are imposed for both the fluid and
the solute at the top and bottom horizontal boundaries, whereas periodicity is applied
at the side boundaries (along $x$).
In dimensionless form, these boundary conditions read as:
\begin{equation}
w^*=0\quad,\quad \frac{\partial C^*}{\partial z^*}=0 \quad\textnormal{ for }\quad z^*=0\quad\textnormal{ and }\quad z^*=\ra\text{  .}
\label{eq:rt_eqbcad1}
\end{equation}
The set of equations~\eqref{eq:rt_eq5a}-\eqref{eq:rt_eq6} has been discretized numerically using a Chebyshev-Tau method.
We refer to \cite{depaoli2016influence,depaoli2016solute,depaoli2019prf,depaoli2019universal} for further details on the flow solver.

\section{Results}\label{sec:rt_resul}
The comparison of the experimental results against numerical simulations is focused on the evolution of the mixing length and on the analysis of the wavenumber power spectra. These spectra are obtained by making a Fourier transform of the instantaneous concentration field and provide a robust characterization of the finger size, as this size can be associated to the wavenumber at which the spectra exhibit a peak.
We consider a wide range of Rayleigh numbers: $4.39\times 10^3\le \ra \le 5.43\times 10^4$.
We refer to Tab.~\ref{tab:listex} for the experimental mass fraction and density difference  values used to obtain these $\ra$.
Above the upper value $\ra_{max} = 5.43\times 10^4$, the density difference between the KMnO$_4$-rich solution (heavy fluid) and water (light fluid) induces large vertical velocities, and the flow is affected by non-Darcy effects \citep{letelier2019perturbative,depaoli2020jfm}.
For lower values of $\Delta\rho $, in contrast, the initial condition is hard to achieve and the duration of the experiments increases, leading to higher chances of inducing density perturbations in the cell from the side and horizontal channels (L, R, T, B in Fig.\ref{fig:sk}).
In the following, we describe the experimental results in terms of flow phenomenology (Sec.~\ref{sec:rt_resu_gen1}), time evolution of the mixing length (Sec.~\ref{sec:rt_resu_gen2}) and wavenumbers (Sec.~\ref{sec:rt_resu_gen3}),
comparing vis-\'a-vis experimental observations against numerical simulations \citep{depaoli2019prf,depaoli2019universal}.
The datasets analyzed in the present study have been deposited in \citep{dataset}.

\subsection{Flow phenomenology}\label{sec:rt_resu_gen1}
 
\begin{figure}
  \center
\includegraphics[width=0.98\columnwidth]{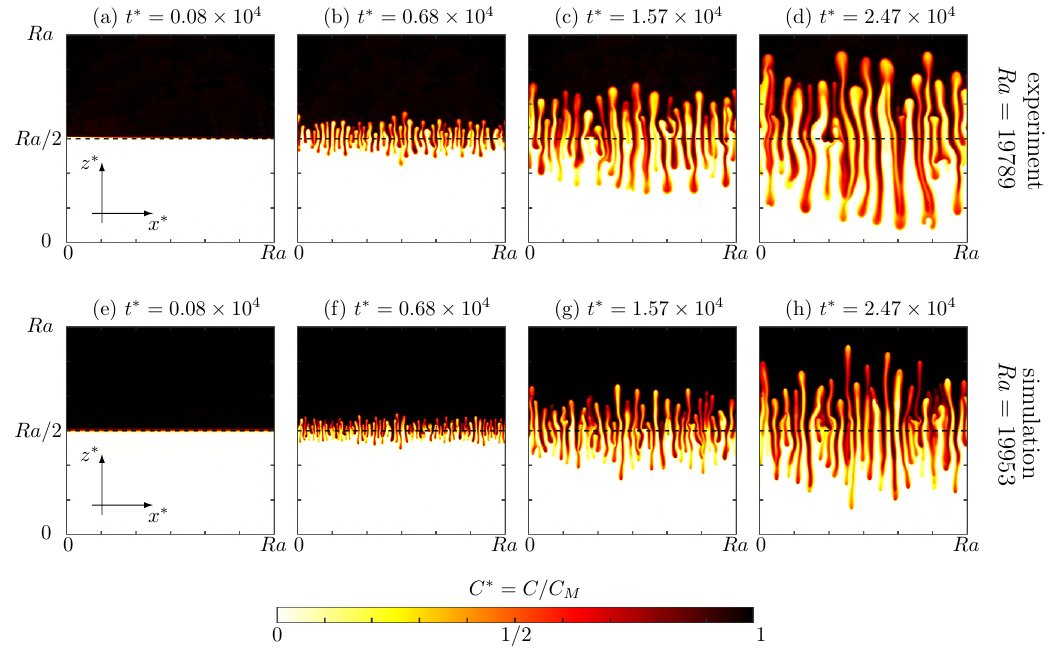}  
\caption{\label{fig:rt_snap2} 
Snapshots of the dimensionless concentration field obtained experimentally (a-d) and numerically (e-h).
The experiment (E4 in Tab.~\ref{tab:listex}, $\ra=19789$) is compared vis-\'a-vis against numerical simulations performed at $\ra=19953$ by \citet{depaoli2019prf}.
The dimensionless time instants ($t^*$) at which the fields refer are indicated
on top of each panel.
We remark that the experiments are realised in a square geometry, whereas the simulations are performed in wider domains (aspect ratio $\pi/2$). 
Therefore, for better comparison, only a portion of the numerical domain having unitary aspect ratio is reported.
The desired initial position of the interface (located at $z=H/2$, i.e. $z^*=\ra/2$) is also shown as a dashed black line.
See also the movie, Movie~S1, in the Supplemental Material \citep{movie_s1} for a comparison of the time-dependent evolution of the concentration fields in the
experiment and in the simulation.
}
\end{figure}

To analyze in detail the flow phenomenology, we report in Fig.~\ref{fig:rt_snap2} the time-dependent dimensionless concentration field ($C^*$) obtained experimentally (\ref{fig:rt_snap2}a-d) and numerically  \cite{depaoli2019prf} (\ref{fig:rt_snap2}e-h).
The experiment (E4 in Tab.~\ref{tab:listex}, $\ra=19789$) is compared against numerical simulations performed at similar Rayleigh number, $\ra=19953$: We consider this value because it is large enough to observe all the main phases of the mixing process for
a significant amount of time.
On top of each panel, the dimensionless time instants ($t^*$) at which the fields are taken are provided.

The concentration field is initially in an unstable configuration (Fig.~\ref{fig:rt_snap2}a,e) with the heavy fluid characterised by the dimensionless concentration $C^*=1$ lying on top of the lighter one ($C^*=0$).
The initial position of the interface, indicated here with the dashed line at $z^*=\ra/2$, is initially flat and horizontal. 
Achieving such a configuration in the experiment is very difficult, and some discrepancy, though not apparent, always exists between the actual fluid-fluid
interface and the desired interface.
This is visible for instance in Fig.~\ref{fig:rt_snap2}(a), where the upper fluid layer seems to be slightly inclined with respect to the horizontal dashed line.
In the experiments, the flow is initially triggered by perturbations of the local density and velocity fields. 
Indeed, to set the initial flow configuration and generate the fluid-fluid interface, some fluid is removed from the side channels and simultaneously injected from the channels at the horizontal boundaries (see Sec.~\ref{sec:calib} for a thorough description of this process).
As a result, the fluid will not be initially still and will be partially affected by the flow circulations induced by the procedure adopted to set the initial condition.
The situation is completely controllable in the simulations, where the interface starts as perfectly flat and is made unstable by adding a perturbation directly to the concentration field (Fig.~\ref{fig:rt_snap2}e). 
The amplitude of the perturbation considered, however, is crucial in determining the onset of the instability \citep{depaoli2019universal}, whereas the late stage dynamics are not sensitive to this parameter. 
In the present study, we considered simulations in which the perturbations applied to the dimensionless concentration field have amplitude $10^{-3}$.

The flow phenomenology associated with the mixing process can be summarised as follows.
The mechanism of diffusion is initially responsible for the flow evolution (Fig.~\ref{fig:rt_snap2}a,e). 
Afterwards, the interfacial layer thickens and, due to the unstable density profile, finger-like structures form.
These fingers are controlled by the interplay of convection, which makes them grow vertically, and diffusion, which reduces the horizontal concentration gradients and promotes mixing \citep{gopalakrishnan2017relative}. 
During this intermediate stage (Fig.~\ref{fig:rt_snap2}b-c,f-g), the fingers grow symmetrically with respect to the domain centerline, and eventually merge.
Indeed, we observe that the number of fingers reduces from $t^*=0.68\times10^4$ (Fig.~\ref{fig:rt_snap2}b,f) to $t^*=1.57\times10^4$ (Fig.~\ref{fig:rt_snap2}c,g). 
We anticipate here that this dynamics is quantitatively confirmed by the wavenumber measurements reported in Sec.~\ref{sec:rt_resu_gen3}.
At a later stage (Fig.~\ref{fig:rt_snap2}d,h), the number of fingers does not increase further.
Eventually, the strength of convection diminishes and the flow evolution is controlled again by diffusion. In this phase (not shown here), the system has achieved a stable configuration in which the average concentration in the lower layer is higher than in the upper layer \citep{depaoli2019prf}. 
This regime, usually referred to as {\it shutdown of convection} in view of the reduced strength of the driving forces \citep{hewitt2013convective,depaoli2016solute}, starts after the tips of the fingers reach the horizontal walls and the effect of the boundaries propagates back to the domain centerline.
The time required to reach the shutdown is large (e.g., of the order of 1 hour for the lower $\ra$ considered here), and hence hard to achieve experimentally because the flow conditions have to be carefully controlled for a considerable amount of time.

It is apparent that the flow evolution observed in the experiments is qualitatively similar to that of numerical simulations, albeit with some differences.
The growth of the fingers, expressed in terms of their tip-to-rear extension, is similar within the same time interval (Fig.~\ref{fig:rt_snap2}b,f and Fig.~\ref{fig:rt_snap2}c,g).
We also refer the reader to the movie, Movie~S1, in the Supplemental Material \citep{movie_s1} for a comparison of the time-dependent evolution of the concentration fields in experiments and simulations. 
However, we also observe a different number of fingers in the two cases (see~Sec.~\ref{sec:rt_resu_gen3} for further discussion).

\subsection{Mixing length}\label{sec:rt_resu_gen2}

The evolution of a Rayleigh-Taylor system can be conveniently analysed by looking at the instantaneous vertical extension of the fluid-fluid interface \citep{boffetta2017incompressible,dalziel1999self}. 
This quantity, defined as mixing length $h$, can be also interpreted as the distance between the upwarding and downwarding tips of the fingers. 
To compute the mixing length in a quantitative way, one can use either global \citep{cook2004mixing} (i.e. integral) or local \citep{gopalakrishnan2017relative,de2004miscible} flow properties. 
In this work, we characterize the mixing length by means of the local concentration field \citep{de2004miscible,depaoli2019prf}, i.e. $h$ is the portion of the domain where the horizontally-averaged concentration field satisfies the condition  $\varepsilon<\overline{C}(z,t)<1-\varepsilon$, with:
\begin{equation}
\overline{C}(z,t)=\frac{1}{H} \int_0^H \frac{C(x,z,t)}{C_M}\diff x    
\label{eq:ml12}
\end{equation}
the concentration field averaged along the horizontal direction $x$, $\varepsilon=3\times10^{-2}$ a threshold coefficient, and $H$ the domain width (in $x$ direction). 
It follows from this definition that $0\le h\le H$ or, in dimensionless terms using the length scale defined in Eq.~\eqref{eq:rt_ra0exp2}, $0\le h^*\le \ra$.

\begin{figure}
\includegraphics[width=0.8\columnwidth]{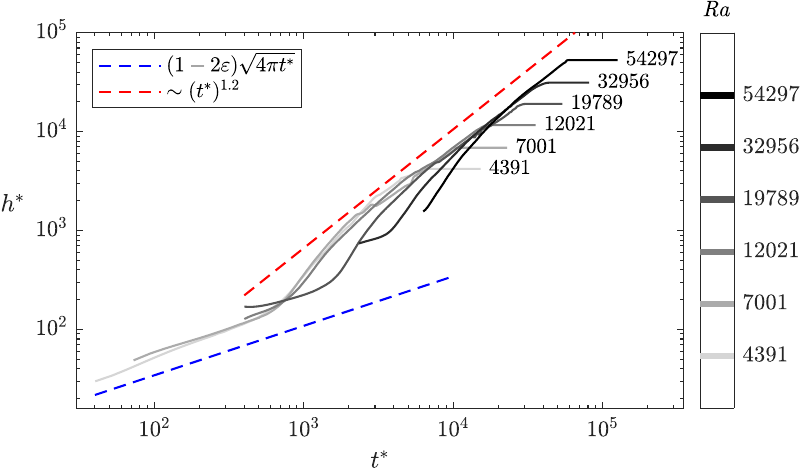}
\caption{\label{fig:rt_diffus} 
Time evolution of the dimensionless mixing length, $h^*(t^*)$ (solid lines), for all the values of the global Rayleigh number considered in this study (and explicitly indicated next to each curve).
The evolution of the system, initially controlled by diffusion, nicely follows the analytical solution, $h^*\approx(1-2\varepsilon)\sqrt{4\pi t^*}$ [Eq.~\eqref{eq:appA3}, dashed blue line]. 
The intermediate phase, dominated by convection, exhibits the anomalous scaling $h^*\sim (t^*)^{1.2}$ first observed by \citet{depaoli2019prf}.
}
\end{figure}

The time evolution of the dimensionless mixing length for all the experiments we performed is shown in Fig.~\ref{fig:rt_diffus}.
As previously observed in simulations in confined domains \citep{depaoli2019prf,depaoli2019universal}, the dynamics consists essentially of three phases in sequence: (i) diffusion-dominated, (ii) convection dominated and (iii) wall-induced shutdown.
During the initial phase, the thickness of the mixing layer is initially controlled by diffusion.
Assuming an initial stepwise concentration profile, the dimensionless concentration field evolves as:
\begin{equation}
C^*(z^*,t^*)=\frac{1}{2}\left[1+\text{erf}{\left(\frac{z^*-\ra/2}{\sqrt{4t^*}}\right)}\right]\text{  ,}
\label{eq:appA2}
\end{equation}
and the corresponding mixing length as:
\begin{equation}
h^*(t^*)=-2\sqrt{4t^*}\text{ erf }^{-1}(2\varepsilon-1)\approx\sqrt{4\pi}(1-2\varepsilon)\sqrt{t^*}
\label{eq:appA3}
\end{equation}
(we refer to \citep{depaoli2019prf} for a detailed derivation).
The same evolution of the mixing length in the initial diffusive regime has been observed in previous numerical works, both in porous media convection \citep{gopalakrishnan2017relative,depaoli2019prf}
and in pure fluids with different initial conditions \citep{Biferale2018}.
Despite the seemingly universal behaviour exhibited by the mixing length, we observe that the initial diffusive stage is hardly captured by the experimental measurements. 
Only the measurements performed at lower $\ra$, namely 4407 and 7023, are in good agreement with the theoretical predictions (blue dashed line in Fig.~\ref{fig:rt_diffus}). 
While the trend nicely follows the predicted behaviour, namely $h^*\sim\sqrt{t^*}$, the magnitude slightly differs from the solution derived in Eq.~\eqref{eq:appA3}. 
This is due to uncertainty of the experimental measurements, which is comparable to the values of $h^*$ at low $\ra$ and early times.
When the fluid-fluid interface is sufficiently thick ($10^3<t<10^4$), it becomes unstable and eventually finger-like structures form. 
When observed via the mixing length, the transition to this convection-dominated regime is smooth, similarly to what has been observed in the simulations.
However, we find that the transition in experiments occurs at different instants depending on the value of $\ra$, as a result of the uncertainty on the initial perturbation of the concentration field, which cannot be reproduced faithfully in the laboratory. 
We also observe that, during the transition, the growth of the mixing length modelled as a power law, i.e. $h^*\sim(t^*)^{\alpha}$, is characterized by an exponent that is higher than that observed asymptotically, in agreement with previous numerical \citep{depaoli2019prf,gopalakrishnan2017relative} and experimental ($\alpha=2$) \citep[]{wooding1969growth} findings.

The fluid mixing is more efficient when convection takes place. 
The formation of finger-like structures marks the beginning of the convection-dominated phase, more persistent in time compared to the initial diffusive stage.
In this phase, fingers grow, merge and finally reach the horizontal boundaries of the cell. 
The measurements reported in this work reveal that the  mixing length grows more than linearly in time, in close agreement with previous simulations (namely, $h^*\sim (t^*)^{1.208}$, \citep{Boffetta2020,depaoli2019prf}).
This scaling is anomalous: It is not expected in two-dimensional porous media convection, where a linear evolution of the mixing length with time is predicted as a result of the balance between buoyancy (i.e., convection) and friction (i.e. dissipation due to viscosity).
The occurrence of a linear growth would imply an asymptotic velocity of the finger tips, which determines also the growth rate of the mixing length, equal to the buoyancy velocity $U$, defined as in Eq.~\eqref{eq:rt_ra0exp2}. We remind the reader that $U$ can be interpreted as the velocity at which the contribution of buoyancy ($g\Delta\rho  k $) balances the dissipation ($\mu$).

In three-dimensional porous media, the growth of the mixing length follows the linear prediction obtained from the physical arguments mentioned above \citep{Boffetta2020}.
The behaviour is different in two dimensions, where the fingers are slenderer compared to their three-dimensional counterpart.
The superlinear evolution of the mixing length we observed, can be ascribed to multiple reasons: (i) the simulated time is not sufficiently large to achieve the asymptotic regime, or (ii) the $\ra$ number of the system is too low, and the effect of diffusion is still dominant.
We speculate that in the present case a combination of these two effects controls the evolution of the flow. 
Initially, the fluid accelerates from zero velocity, and this corresponds to a finite-time effect.
At a later stage, when the initial condition has been absorbed by the system and fluid velocity is non-zero, the finite-size (i.e., finite-$\ra$) effect comes into play, via the domain boundaries.
This non-linear behaviour seems indeed to vanish at higher $\ra$ numbers \citep{boffetta2022dimensional}, but in absence of boundaries.
We remark here that a linear growth of the mixing length has been observed numerically \citep{gopalakrishnan2017relative}, but only in the limit of lower Rayleigh numbers, namely $\ra\approx 4\times10^3$.
For these values of $\ra$, also previous numerical simulations \citep{depaoli2019prf} and present experiments do not exhibit a clear superlinear behaviour. 
In addition, the experimental data reported by \citet{wooding1969growth}, which refer to higher Rayleigh numbers (approximately $10^3\le\ra\le7\times10^4$) are also well fitted by a linear scaling. 
However, it must be pointed out that the data have been collected at very low acquisition rate, making the amount of available measurements very scarce for a time-dependent flow analysis.
We are not able to explain, in this case, why the superlinear scaling was not observed.
One possible reason could be attributed to the method adopted to change the Rayleigh number, consisting of varying the orientation of the cell with respect to gravity.
Another possibility is represented by the absence of horizontal boundaries.
A conclusive answer on the asymptotic scaling of the mixing length could be obtained with the aid of Darcy simulations or experiments at higher $\ra$, ($\approx10^6$), which are beyond the present capabilities.

\begin{figure}
\center
\includegraphics[width=0.65\columnwidth]{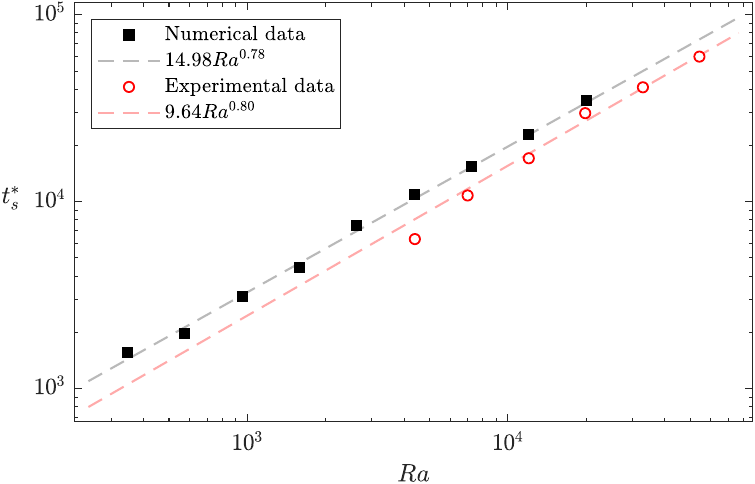}
\caption{\label{fig:rt_bound1} 
Time taken by the fingers to reach the horizontal boundaries $(t_s^*)$ as a function of $\ra$. 
Experimental (present work) and numerical \citep{depaoli2019prf} results are reported as open ({\large $\circ$}) and filled symbols ({\footnotesize $\blacksquare$}), respectively. 
The best-fitting power law for present results (red dashed line) is $9.64\times\ra^{0.80}$, in excellent agreement with numerical findings ($14.98\times\ra^{0.78}$, black dashed line).
Experimental results correspond to the mean values obtained for several flow realisations at fixed $\ra$.
}
\end{figure}

To conclude the discussion on the mixing length, we now examine
the wall-induced convective shutdown phase \citep{depaoli2019prf}, which sets in when the high (resp. low) concentration fingers reach the lower (resp. upper) horizontal boundary.
In this phase, the local concentration gradients decrease, weakening the density differences that drive the flow within the domain.
As a result, convection vanishes and the overall mixing rate, identified for instance by means of the mean scalar dissipation rate, decreases dramatically \citep{depaoli2019universal}.
This effect is crucial for applications of practical interest (e.g., carbon dioxide sequestration in saline aquifers), in which the fluids are confined by impermeable boundaries that play a key role in the solute mixing process. 
A simple way to quantify the convective shutdown is to measure the time $t_s^*$ required for the fingers to reach the horizontal boundaries.
The values of $t_s^*$ obtained as a function of the Rayleigh number are shown in Fig.~\ref{fig:rt_bound1}.
Experimental and numerical \citep{depaoli2019prf} results are reported as open ({\large $\circ$}) and filled symbols ({\footnotesize $\blacksquare$}), respectively. 
The best-fitting power law for present results (red dashed line) is $9.64\times\ra^{0.80}$, in excellent agreement with numerical results ($14.98\times\ra^{0.78}$, black dashed line).
This is not surprising, given the close correspondence observed for the evolution of the mixing length.
In leading order, we can estimate $t^*_s= t^*:h^*(t^*)=\ra$ which gives $t^*_s\sim\ra^{1/1.2}\sim\ra^{0.83}$, consistent with the $\ra^{0.80}$ scaling provided by data fitting.

Despite the excellent agreement of the scaling exponents, we also
observe some discrepancies between simulations and experiments. 
In particular, for a given $\ra$, the value of $t^*_s$ is lower in the experiments, leading to 
a lower value of the multiplicative coefficient obtained from data fitting. 
We attribute this discrepancy to the perturbation of the interfacial concentration field.
In the experiments, the amplitude of this perturbation is hard to control, and it is likely to be  larger compared to the numerical simulations considered, leading to an earlier onset of convective instabilities.
Another limitation of our experimental setup is represented by the range of Rayleigh numbers that can be investigated.
To be able to achieve a Darcy flow and to control effectively the boundary conditions, the values of domain size ($H^*$) and the density difference ($\Delta\rho $) cannot be too large (see also  Sec.~\ref{sec:expset}).
The values of $\ra$ considered here are sufficiently large to overlap with and go beyond (up to 2.5 times, approximately) the largest Rayleigh numbers simulated numerically.
On the other hand, achieving smaller values of $\ra$ is also challenging. 
Lower values of $\Delta\rho $ correspond to lower velocities, and therefore longer experiments. Smaller values of $H^*$, i.e. shorter domains, would instead required more careful setup of the initial flow configuration. 

The findings discussed in this section prove that, in terms of evolution of the mixing length, two-dimensional Darcy simulations and Hele\hyp{}Shaw experiments produce similar results when the same $\ra$ is considered. 
However, the ability of the experimental measurements to capture the initial development of the flow is lower due to the setup used, and numerical simulations should be considered as more reliable to investigate this stage of the flow evolution.

\subsection{Wavenumbers}\label{sec:rt_resu_gen3}

\begin{figure}[t!]
\center
\includegraphics[width=0.95\columnwidth]{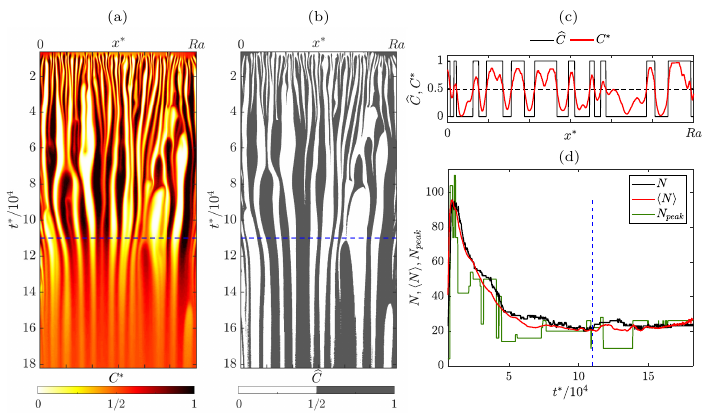}
\caption{\label{fig:wn1}
Measurements of the number of fingers over time for $\ra=5.43\times 10^4$ (E6 in Tab.~\ref{tab:listex}).
Space-time map, consisting of the evolution of the concentration field along the centerline, $C^*(x^*,z^*=\ra/2,t^*)$, is shown in (a). 
To compute the finger number, the space-time map is binarized, $\widehat{C}(x^*,z^*=\ra/2,t^*)$, choosing $C^*=1/2$ as threshold (b).
The concentration profile (red line) and the discretized profile (black line) measured along the centerline at $t^*\approx1.1\times10^5$ (indicated by the dashed blue line in panels a, b and d) are also reported as a function of the horizontal coordinate $x^*$ (c).
Finally (panel~d), the time-dependent finger number is computed as from the binarized field $\widehat{C}$ (black line), from power-averaged mean wavenumber as defined in Eq.~\eqref{eq:def_kavg} ($  \langle N \rangle =L\langle k \rangle/\pi$, red line) and from a $k_\text{peak}$ (green line), i.e. the wavenumber maximizing the spectrum amplitude of $C$.
See also the movie, Movie~S2, in the Supplemental Material \citep{movie_s2} for the time-dependent evolution of the wavenumber and the corresponding concentration field.}
\end{figure}

The flow dynamics is complex and characterised by a continuous change of the fluid-fluid interface position. 
To provide a quantitative estimate of the flow topology evolution, we analyze here the number of fingers and the wavenumbers at the centerline of the domain ($z^*=H^*/2$). 
The process of flow analysis is illustrated in Fig.~\ref{fig:wn1}, where an experiment at $\ra=5.43\times 10^4$ is considered for reference. 
First, we build the space time map at the centerline, where the interface is initially located, i.e. we report in Fig.~\ref{fig:wn1}(a) the contour of the time-dependent dimensionless concentration field, $C^*(x^*,z^*=\ra/2,t^*)$. 
The simplest way to define the number of structures in the system is to count the upwarding and downwarding fingers. 
To this aim, the spacetime map is first binarised [$\widehat{C}(x^*,z^*=\ra/2,t^*)$, Fig.~\ref{fig:wn1}(b)] with respect to a threshold value corresponding to the mean concentration ($C^*=1/2$), and then the number of fingers is counted [Fig.~\ref{fig:wn1}(c)].
The process is iterated during the entire flow evolution, and the number of fingers, $N(t)$, is computed as the number of regions with different binarized concentration values [Fig.~\ref{fig:wn1}(d), black line]. 
Due to noise in the experimental measurements, spurious fluctuations of the concentration field may occur. To avoid counting these as fingers, we discarded contiguous regions of the binarized map that are smaller than 4 pixel (corresponding to a physical length of 350~$\mu$m).
We observe in Fig.~\ref{fig:wn1}(d) that this threshold is extremely sensitive to variations of the concentration field, i.e. it appears to fluctuate without an apparent physical reason.  
A more robust characterization of the flow topology is provided by the wavenumbers associated with the power spectra of the concentration field. 
We employ a Fourier decomposition of the concentration profile at the cell centerline to estimate the number of fingers, and compute the power-averaged mean wavenumber at each time instant, defined as \citep{de2004miscible}:
\begin{equation}
\langle k \rangle = \frac{\int k_n {|C_n(t^*)|}^2 \text{d}k_n}{\int {|C_n(t^*)|}^2 \text{d}k_n}\text{ , }
\label{eq:def_kavg}
\end{equation}
with $k_n$ the wavenumber and  ${|C_n(t^*)|}^2$ the squared amplitude of the $n$-th basic harmonic function obtained from the Fourier decomposition of the fluctuation of $C^*(x^*,z^*=1/2,t^*$) about its mean value. 
We observe that in this case [red line in Fig.~\ref{fig:wn1}(d)] the evolution of the number of fingers, $\langle N\rangle = L\langle k \rangle /\pi$, is smoother compared to the simple finger count, $N$, and also compared to the peak number, defined as $N_\text{peak}=L k_\text{peak}/\pi$ [orange line in Fig.~\ref{fig:wn1}(d)], where $k_\text{peak}$ is the wavenumber value for which $|C_n(t^*)|$ is maximum.
Therefore, in the following, we will focus the analysis considering the power-averaged mean finger number $\langle N \rangle$ only, which appears similar to but less fluctuating of the finger number $N$.

\begin{figure}[t!]
\center
\includegraphics[width=0.49\columnwidth]{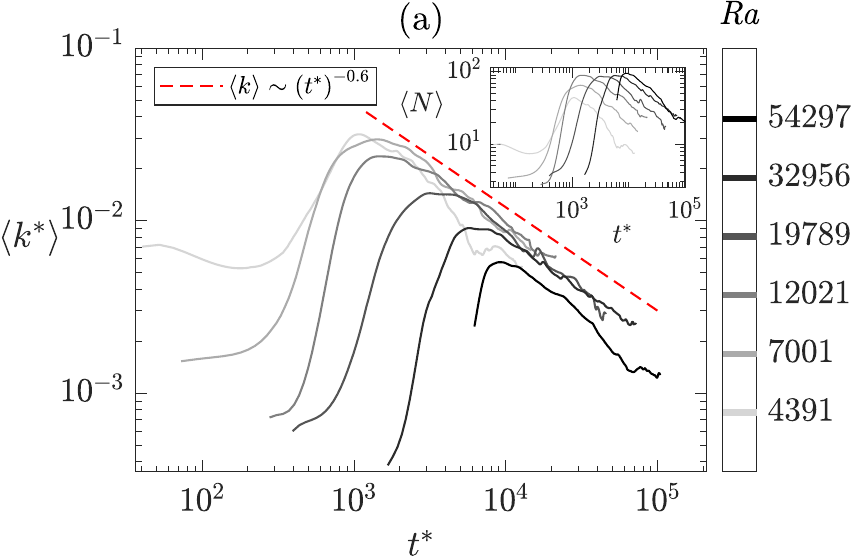}
\includegraphics[width=0.49\columnwidth]{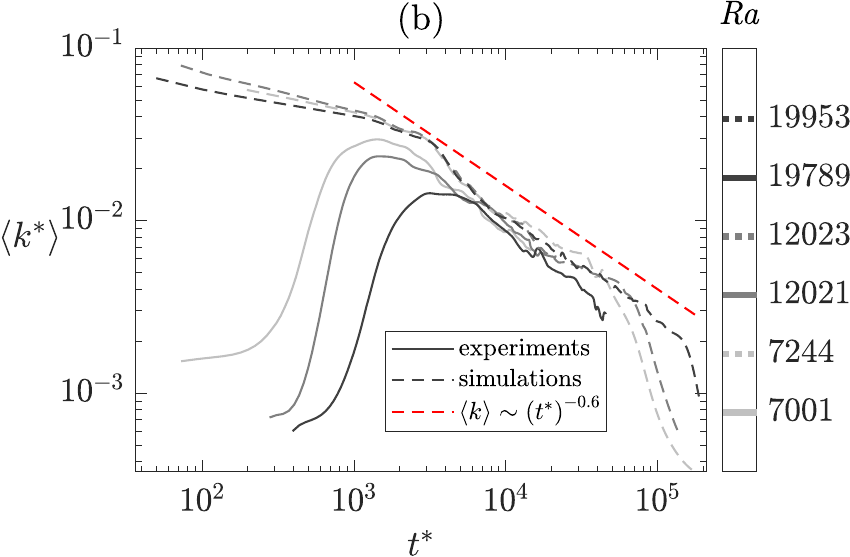}
\caption{\label{fig:wn2}
(a)~Experimental measurements. 
The evolution of the power-averaged mean wavenumber, $\langle k \rangle$ (defined in Eq.~\eqref{eq:def_kavg}, main panel), and the corresponding finger number $\langle N \rangle$ (inset).
(b)~Comparison of experimental (solid lines) and numerical (\citep{depaoli2019universal}, dashed lines) measurements. For ease of reading, three values of Rayleigh number are used to compare the behaviour of $\langle k \rangle$.
}
\end{figure}

A comparison of the number of fingers measured in the experiments is reported in the inset of Fig.~\ref{fig:wn2}(a).
In general, all the experiments are characterized by a similar dynamics.
The number of fingers, initially low, increases up to a maximum value. 
Afterwards, the fingers start to merge and increase in size. This phase, which is also visible from the spacetime maps in Fig.~\ref{fig:wn1}(a-b), may start earlier at low $\ra$ and later at high $\ra$,
spanning a relatively long interval ($10^3\le t^* \le 10^4$) in Fig.~\ref{fig:wn2}(a).
During growth, $\langle N \rangle$ appears to decrease in time at the same rate for all $\ra$ considered.
This universal behaviour is more apparent when the results are compared in terms of mean wavenumber $\langle k \rangle$ (main panel of Fig.~\ref{fig:wn2}a). 
After entering the convection-dominated regime, all curves exhibit the same trend.
A first observation is that the merging rate of the fingers is in agreement with the scaling $\langle k \rangle\sim (t^*)^{-0.6}$ reported in numerical simulations \citep{depaoli2019universal}. 
In contrast, the behaviour is different during the diffusive phase, when the wavenumber is clearly dependent on $\ra$.

In Fig.~\ref{fig:wn2}(b), we compare present results (solid lines) against previous numerical findings \citep{depaoli2019universal} (dashed lines).
For ease of reading, only three values of $\ra$ number are considered. 
Two differences are apparent: 
(i)~unlike experiments, simulations \citep{depaoli2019universal} have been performed on domains of constant dimensionless width (namely, 19953), and 
(ii)~to make reliable comparisons among different Rayleigh numbers, the same initial perturbation has been employed, thus making the early stage flow evolution also universal and independent of $\ra$.
This highlights again a possible limitation of the experimental procedure employed, which makes it hard to precisely control the initial flow conditions.
As a result, the initial phase is characterised by large discrepancies between numerical and experimental results (Fig.~\ref{fig:wn2}b).
In the late stage of the flow dynamics, numerical and experimental measurements are in fair agreement.
As a side note, we remark that present results, shown in terms of wavenumbers associated with the concentration field, are compare against the simulations of \citet{depaoli2019universal}, where the power spectra where obtained from the scalar dissipation fields. 
However, even the experimentally-measured merging rate is well described by the same scaling law, $t^{-0.6}$.
Present results are in fair agreement with previous experimental findings \citep{wooding1969growth,gopalakrishnan2017relative}, in which the wavenumber is observed to evolve according to $(t^*)^{-1/2}$. 
This scaling has been also derived analytically \citep{wooding1969growth,slim2014solutal} assuming that the fingers are slender, so that vertical derivatives (e.g., diffusion of solute) are negligible compared to the horizontal ones.
In the frame of this model, the flow is driven by vertical advection, whereas lateral diffusion acts as dissipative mechanism.
We refer to \citet{slim2014solutal} for a detailed derivation of this result.
The slight difference in the scaling exponent, compared to the analytical prediction, maybe attributed to the simplified assumptions considered in the model derivation.
The scaling $\langle k \rangle \sim (t^*)^{-1/2}$ has been also suggested by \citet{gopalakrishnan2017relative}, who interpreted this evolution as a diffusion-dominated finger growth, by analyzing the relative contribution of convective and diffusive terms within a finger (but away from the finger tips).

\section{Conclusions}\label{sec:rt_concl}
In this work, we analyzed with the aid of high-resolution experiments the Rayleigh-Taylor instability in a Hele\hyp{}Shaw flow.
The system consists of two miscible fluid layers, namely water and an aqueous solution of KMnO$_4$, having different density but same viscosity, placed in an initial unstable configuration. 
The density difference is induced by the presence of the solute (KMnO4), which produces also a colour gradient by which we are able to infer the concentration field.
The system is fully described by the Rayleigh number, $\ra$, which measures the relative strength of convective and dissipative contributions. 
Present experimental results have been performed taking great care to achieve a proper initial condition and minimizing the local perturbations of the concentration and velocity fields, which trigger the growth of interfacial finger-like instabilities.
In addition, a high resolution in space and time has been achieved by considering relatively large domains, and an accurate reconstruction of the concentration field is performed. 
In this respect, present experimental results go beyond previous measurements published in archival literature.

We compared present findings against previous numerical measurements in two\hyp{}dimensional and confined media to which the Darcy equations can be successfully applied, as for instance porous media. \citep{depaoli2019universal,depaoli2019prf}.
The dynamics of the flow is first controlled by diffusion, which is responsible for the thickening of the interface, initially flat and horizontal. 
Afterwards, the flow evolution is controlled by convection.
The unstable fluid-fluid density front promotes the formation of high-wavenumber instabilities, which grow and merge into large and more persistent fingers.
Finally, due to the combined action of convective mixing in a confined domain and lateral diffusion across the fingers interface, the solute distribution becomes uniform within the porous slab, with a corresponding stable concentration profile.
During this stage, referred to as wall-induced convective shutdown, the flow is again controlled by diffusion.
The flow evolution is quantified and compared against theoretical and numerical predictions with the aid of two main observables: the mixing length, $h^*(t)$, and the power-averaged mean wavenumber, $\langle k \rangle$.

Initially, the evolution of the mixing length is in fair agreement with the theoretical predictions obtained for flows in the absence of convection ($h^*(t)\sim (t^*)^{1/2}$).  
In the intermediate regime, in which convection dominates, $h^*(t)$ is observed to grow with a superlinear scaling, $h^*(t)\sim (t^*)^{1.2}$, confirming previous numerical findings \citep{depaoli2019prf,Boffetta2020}. 
This scaling is anomalous, as it differs from the theoretical prediction, $h^*(t)\sim t^*$, possibly as a result of a finite size (i.e., finite $\ra$) effects.
To further compare present experimental findings against previous numerical simulations, we compute the time $t_s^*$ taken for the fingers to reach the horizontal boundaries of the domain.
Also in this case, experiments ($t_s^*\sim\ra^{0.80}$) are in excellent agreement with the simulations ($t_s^*\sim\ra^{0.78}$).

Finally, we analysed the flow dynamics by looking at the finger distribution at the domain centerline.
The evolution of the characteristic finger size has been quantified by means of the power-averaged mean wavenumber, $\langle k \rangle$.
During the convection-dominated phase, the centerline finger dynamics exhibits a universal behaviour that is independent of $\ra$.
Also in this case, we found an excellent agreement with scaling laws derived in numerical simulations \citep{depaoli2019universal}, namely $\langle k \rangle\sim (t^*)^{-0.6}$.

The analysis carried out in this paper shows that Hele\hyp{}Shaw experiments can faithfully reproduce the dynamics of convective Rayleigh-Taylor instabilities in two-dimensional and confined porous media.
However, it also shows that the proposed experimental approach presents some important limitations, which call for future improvements.
First off, the desired initial fluid-fluid interface consists of a thin, flat and horizontal contact line, which is experimentally challenging to achieve.
In addition, the instability is initially triggered by local perturbations of the concentration and velocity fields, and hence hard to control.
Finally, experiments at low $\ra$ are characterised by low values of vertical velocity, which slow down the mixing process and make the experiments to last longer.
As a result, the experimental conditions have to be monitored for longer times, increasing the difficulty of the experimental procedure.

\section*{Acknowledgments}
This research was funded in part by the Austrian Science Fund (FWF) [Grant J-4612].
For the purpose of Open Access, the authors have applied a CC BY public copyright license to any Author Accepted Manuscript (AAM) version arising from this submission. 
D.P. gratefully acknowledges the financial support provided from Project No. 2020-1-IT02-KA103-078111 funded by UE Erasmus.
Mr. Werner Jandl and Mr. Franz Neuwirth are gratefully acknowledged for their help with the experimental work.
Eliza Coliban and Mobin Alipour are also acknowledged for their contribution during the initial stage of this work.
We are grateful to the anonymous referees for the comments provided.

\appendix
\section{Additional details on fluid properties}\label{sec:appfluids}

\begin{figure}
  \centering
\includegraphics[width=0.85\columnwidth]{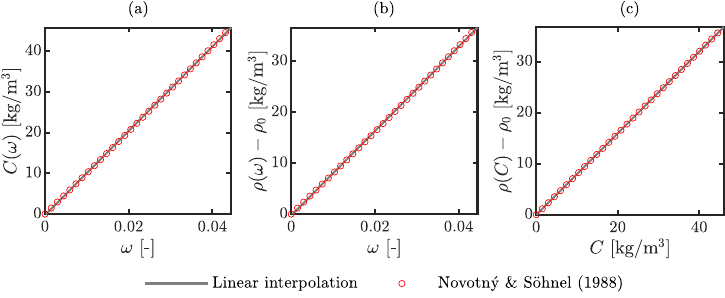}
\caption{\label{fig:fl}
Given the definition~\eqref{eq:defmf} and the empirical correlations~\eqref{eq:novot} \citep{Novotny1988}, the relative dependence of density of the solution, $\rho$, mass fraction, $\omega$, and solute concentration, $C$, is obtained.
We report here $C(\omega)$~(panel~a), $\rho(\omega)-\rho_0$~(panel~b) and $\rho(C)-\rho_0$~(panel~c), being $\rho_0$ the water density.
The profiles shown here correspond to $\vartheta=25^{\circ} C$.
We observe that the empirical values (symbols) are very well fitted by linear functions (solid lines). 
In the instance of density and concentration (panel~c), the linear function corresponds to Eq.~\eqref{eq:rt_eq4a}.} 
\end{figure}

Following the empirical correlations of \citet{Novotny1988}, the density of an aqueous solution of KMnO$_4$ can be expressed as a function of the solute concentration ($C$) and the temperature of the solution ($\vartheta$) as:
\begin{equation}
\rho(C,\vartheta)=f(\vartheta, C)=\rho_0(\vartheta)+ A_1(\vartheta)C+A_2(\vartheta)C^{3/2}\text{  ,} 
    \label{eq:novot}
\end{equation}
where the water density $\rho_0(\vartheta)$ and the temperature-dependent coefficients $A_1(\vartheta)$ and $A_2(\vartheta)$, are given by:
\begin{align}
    \rho_0(\vartheta) &= +9.997\times 10^2+2.044\times 10^{-1}\cdot \vartheta-6.174\times 10^{-2}\cdot \vartheta^{3/2}\text{  ,} \\
    A_1(\vartheta) &= +1.223\times 10^{2}-1.029\times 10^{-1}\cdot \vartheta+8.093\times 10^{-3}\cdot \vartheta^2\text{  ,} \\
    A_2(\vartheta) &= -1.485\times 10^{1}+9.079\times 10^{-1}\cdot \vartheta-7.566\times 10^{-3}\cdot \vartheta^2\text{  .} 
\end{align}
Provided that the mass fraction of the solution is defined as: 
\begin{equation}
\omega(C) = \frac{C}{\rho(C)}\text{  ,} 
    \label{eq:defmf}
\end{equation}
it is possible to determine the respective dependency of density, mass fraction and concentration, which we report in Fig.~\ref{fig:fl}. 
The density of the mixture, $\rho(C)$ is well approximated by a linear function of the solute concentration [Eq.~\eqref{eq:rt_eq4a}], as can be seen in Fig.~\ref{fig:fl}c).

\vspace{-0.6cm}

\section{Measurements calibration}\label{sec:appmeas}

\begin{figure}
    \centering
    \includegraphics[width=0.65\columnwidth]{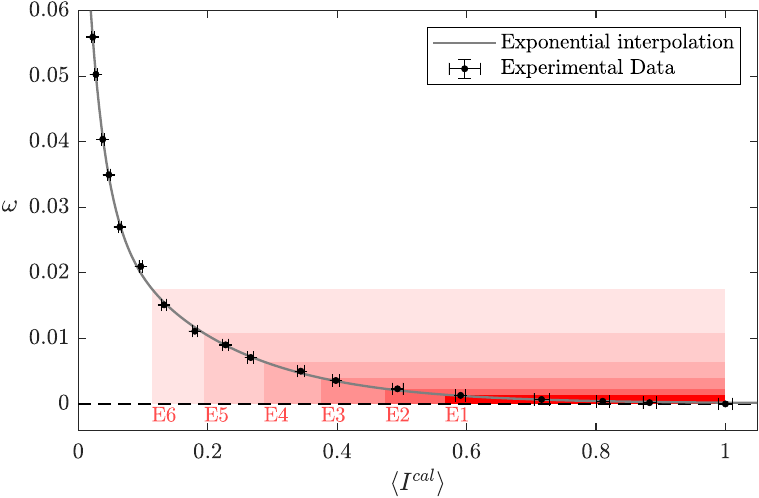}
    \caption{\label{fig:cal_interp}
    Example of calibration curve for a given illumination condition. 
    A number of samples having different mass fraction ($\omega$) are used to uniformly fill the cell. 
    The average value of light intensity, $\langle I^{cal} \rangle$, is used to build the curve.
    Experimental measurements (symbols) are well fitted by a combination of exponential functions, $\omega(\langle I^{cal} \rangle)=a_0+a_1\exp(b_1\langle I^{cal} \rangle)+a_2\exp(b_2\langle I^{cal} \rangle)$ (solid line).
    Uncertainties on the light intensity and mass fraction are also shown (see Sec.~\ref{sec:expset} for further details). 
    Finally, the mass fraction-light intensity space considered in each experiment, labelled as in Tab.~\ref{tab:listex}, is highlighted in red.
    The measurements are performed where the sensitivity of the method employed is higher, i.e. for low values of mass fraction.
    }
\end{figure}

When the two-dimensional distribution of light intensities is determined, each value of light intensity is assigned a value of mass fraction, $\omega(x,z)$.
This step is achieved with the aid of calibration curves, an example of which is reported in Fig.~\ref{fig:cal_interp}. 
For a given illumination condition, the cell is filled with solution with a known value of mass fraction, $\omega$, and the average light intensity over the cell, $\langle I^{cal}\rangle$, is recorded.
The higher the concentration of KMnO$_4$, the lower the transmitted light intensity \citep[Lambert-Beer law,][]{ingle1988spectrochemical}. 
Experimental measurements (symbols) are well fitted by a combination of exponential functions, $\omega(\langle I^{cal} \rangle)=a_0+a_1\exp(b_1\langle I^{cal} \rangle)+a_2\exp(b_2\langle I^{cal} \rangle)$ (solid line).
Further details on the process of calibration are available in \citep{alipour2019PAMM,depaoli2020jfm,alipour2020concentration}.

\section{Additional details on the initial condition}\label{sec:appincond}

\begin{figure}
    \centering
\includegraphics[width=0.99\columnwidth]{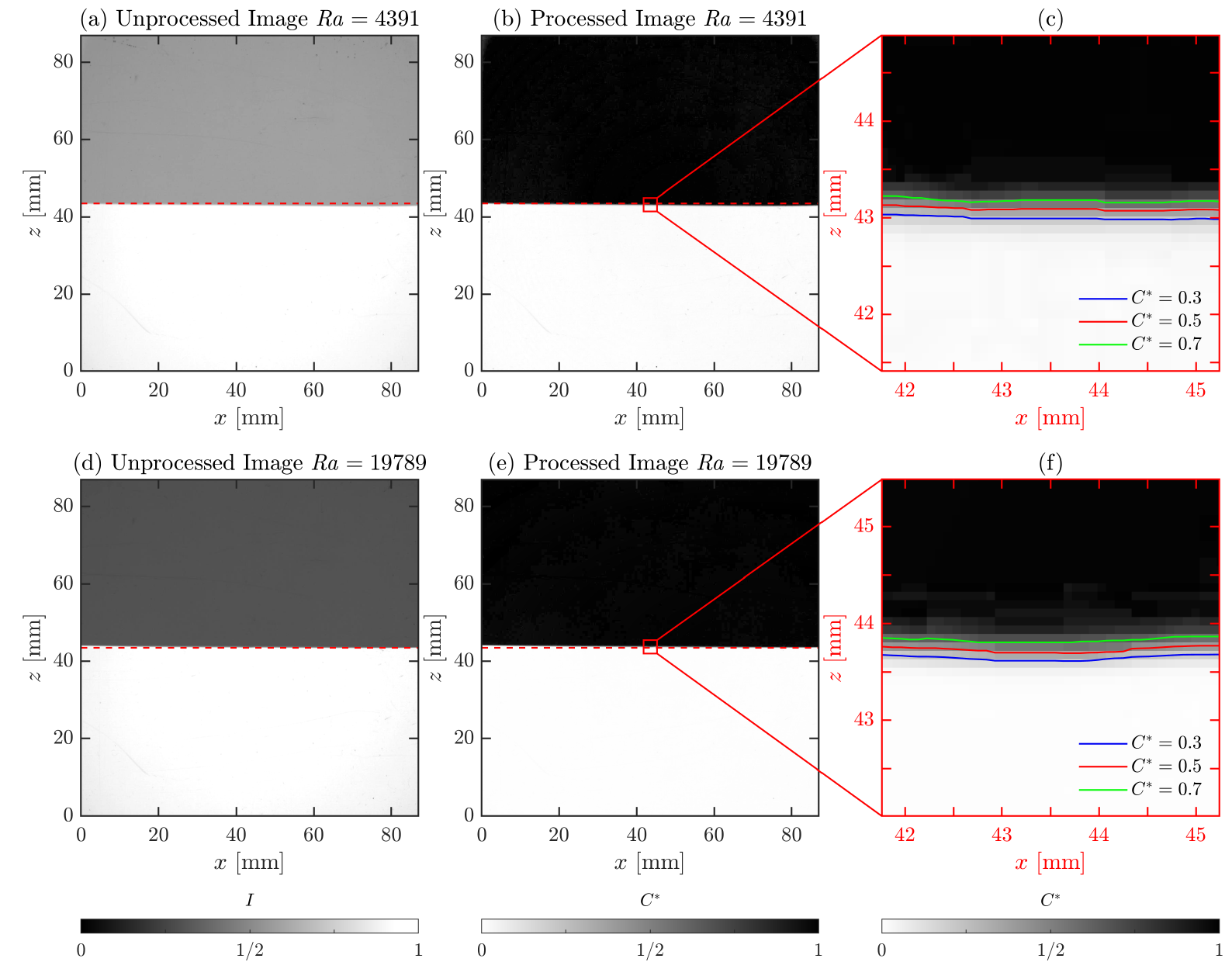}    
    \caption{\label{fig:interf}
    Initial conditions for two different experiments, namely E1 and E4, characterized by $Ra = 4391$ (panels a, b and c) and $Ra = 19789$ (panels d, e and f), respectively. 
    The domain centerline (red dashed line) is indicated on the unprocessed light intensity fields $I$ (panels a, d) and on the normalized inferred concentration fields $C^*$ (panels b,e). 
     In panels (c, f), three iso-contours of dimensionless concentration ($C^*= 0.3$ - blue, $C^*= 0.5$ - red, and  $C^*= 0.7$ - green) are used to identify the interface shape in a close up view of the interfacial region.
     We observe that at both macroscopic (panels~a, b, d, e) and interfacial (panels~c, f) levels, the fluid-fluid separation region is well defined within a narrow area.
    }
\end{figure}

Large values of solute concentration ($C_M$) produce strong concentration gradients across the initial fluid-fluid interface, making a sharp separation between the two fluids hard to obtain.  
Therefore, to achieve a well-defined initial interface up to the $Ra$ values considered here, the flow rate provided to the cell is increased with $\ra$. 
In Fig.~\ref{fig:interf}, the initial condition is shown in detail for the experiments E1 and E4 (namely at $Ra = 4391$ and $Ra = 19789$), and for both the unprocessed light intensity field, $I$, (Fig.~\ref{fig:interf}a,d) and the inferred dimensionless concentration field, $C^*$ (Figs.~\ref{fig:interf}b,e). 
We observe that the experimental procedure employed is suitable to obtain a macroscopic sharp and straight interface in correspondence of the domain centerline, represented by the dashed red line in Figs.~\ref{fig:interf}(a,b,d,e). 
Further information on the local interface shape can be obtained from the close up view of the interfacial region, shown in Figs.~\ref{fig:interf}(c,f), where the iso-contours at $C^* = 0.3$, $C^* = 0.5$ and $C^* = 0.7$ (same as in Fig.~\ref{fig:rec}) are shown.
We observe that, despite the large concentration contrast between the two values of $\ra$ considered, the interface thickness (quantified by the vertical spacial separation between the iso-contours in Figs.~\ref{fig:interf}c,f) is similar, and it is comparable with the spatial resolution of the measurements  (1~pixel $\approx90$~$\mu$m).

\bibliographystyle{abbrvunsrtnat}
\bibliography{bibliography}

\end{document}